\documentclass{elsarticle}
\usepackage{url}
\usepackage{enumitem}
\usepackage{array}
\usepackage{multirow}
\usepackage{mdwlist}

\begin{document}

\begin{frontmatter}

\title{A Methodology of Guiding Web Content Mining and Knowledge Discovery in Evidence-based Software Engineering}

\author{Zheng Li  \\
	Lund University, Lund, Sweden}

\author{\\Yan Liu \\
	Concordia University, Montreal, Canada}

\begin{abstract}

Systematic Literature Review (SLR) is a rigorous methodology applied for Evidence-Based Software Engineering (EBSE) that identify, assess and synthesize the relevant evidence for answering specific research questions. Benefiting from the booming online materials in the era of Web 2.0, the technical Web content starts acting as alternative sources for EBSE. Web knowledge has been investigated and derived from Web content mining and knowledge discovery techniques, however they are still significantly different from reviewing academic literature. Thus the direct adoption of Web knowledge in EBSE lacks of systematic guidelines. In this paper, we propose to make an SLR adaptation to bridge the aforementioned gap along two stages. Firstly, we follow the general logic and procedure of SLR to regulate Web mining activities. Secondly, we substitute and enhance particular SLR processes with Web-mining-friendly methods and approaches.  At the second stage, we mainly focus on adapting Conducting Review by integrating a set of automated components ranging from programmatic searching to various text mining techniques.
\end{abstract}


\end{frontmatter}

\section{Introduction}
Evidence-Based Software Engineering (EBSE) treats empirical primary studies as evidence to investigate software engineering practice, tools and standards \cite{Dyba_Kitchenham_2005}. As the standard and rigorous methodology applied for EBSE, Systematic Literature Review (SLR) has been widely accepted in academia to identify, assess and synthesize the relevant evidence for answering specific research questions \cite{Kitchenham_Charters_2007,Zhang_Babar_2011}. According to the popular aims of implementing a systematic review \cite{Lisboa_Garcia_2010}, the results of an SLR can help identify current research gaps and also provide a solid background for future research activities in a particular field.

The existing SLR implementations collect empirical evidence mostly from the various digital libraries. Given the booming online materials in the era of Web 2.0, the technical Web content becomes an alternative source of empirical evidence in software engineering. Numerous studies have derived results from Web content mining and knowledge discovery to gain evidence of software engineering practices \cite{Bajaj_Pattabiraman_2014,Barua_Thomas_2014,Li_Liang_2013,Lu_Zhu_2013,Vasilescu_Filkov_2013}. Web sites that support the interactions and discussions in the programming question and answer (Q\&A) are often used. For example, popular topics and technology trends in the software engineering community can be revealed through categorizing and analyzing the developer discussion repositories such as Stack Overflow\footnote{\url{http://stackoverflow.com}} \cite{Bajaj_Pattabiraman_2014,Barua_Thomas_2014}. More interestingly, the regular patterns of developer activities and development processes can even be identified \cite{Vasilescu_Filkov_2013}. Furthermore, in addition to those technical websites, industrial forums \cite{Lu_Zhu_2013} and Web media \cite{Li_Liang_2013} have also been employed for empirical evidence aggregation.

\begin{figure}
\centering
\includegraphics{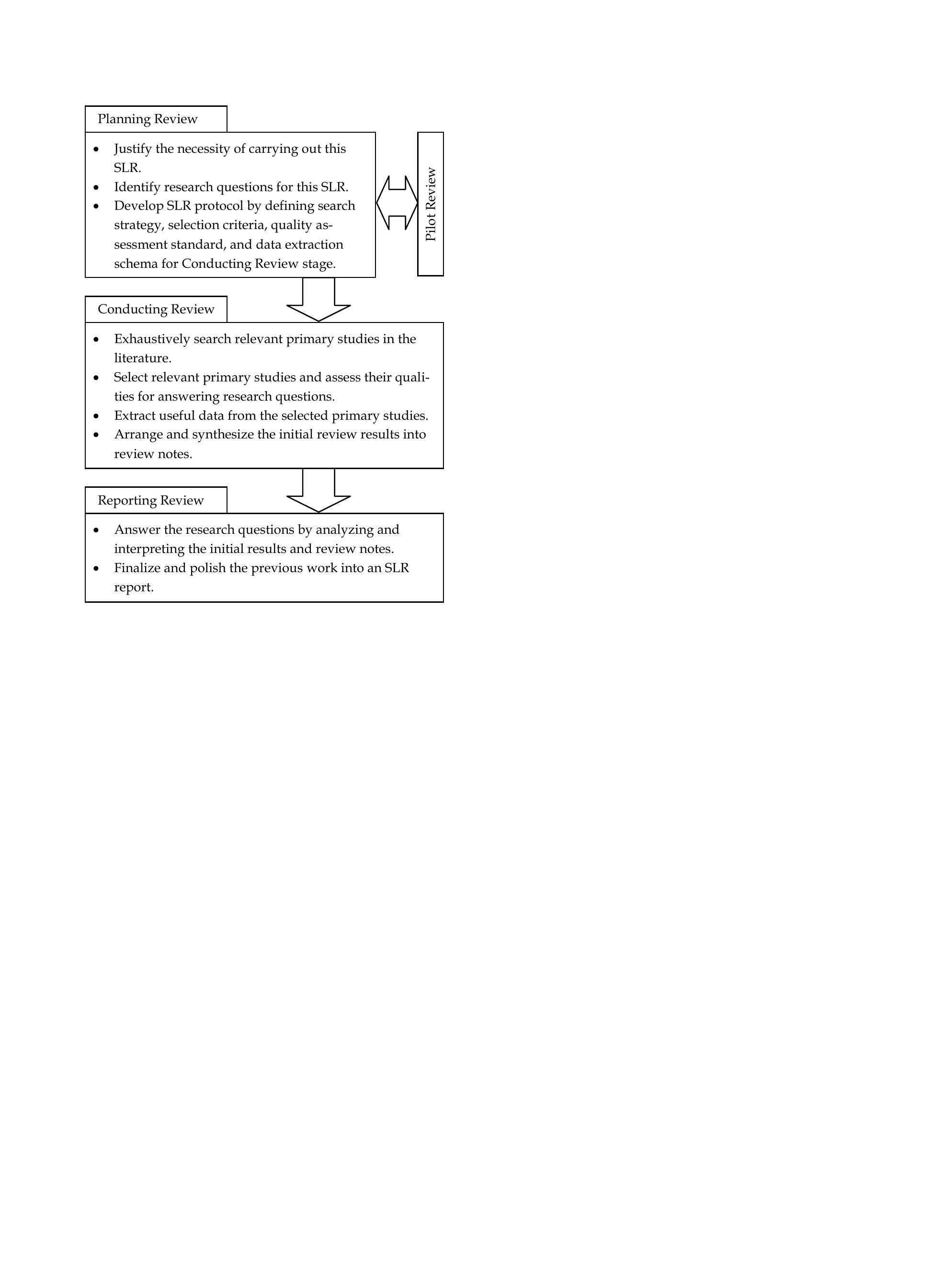}
\caption{Procedure of implementing a systematic literature review.}
\label{fig_SLR_procedure}
\end{figure}

To the best of our knowledge, there is still a lack of a systematic methodology to guide Web content mining and knowledge discovery so that the techniques can be adopted in SLR. In this paper, a methodology refers to ``an organised set of principles which guide action in trying to `manage' (in the broad sense) real-world problem situations" \protect\cite{Checkland_Scholes_1999}. The ``action" in this context indicates a set of necessary activities of Web mining and knowledge discovery. However, most of the relevant studies mainly discussed their findings, while not focusing on the study activities, not to mention the ``principles" to guarantee study objectiveness, traceability or reproducibility. Given the generic procedure of review activities (cf.~Figure \ref{fig_SLR_procedure}), we consider the SLR methodology for mining Web contents. Nevertheless, there are significant differences between the academic literature review and the Web content mining. Here we highlight three of them:

\begin{enumerate}[label=\textit{\arabic{enumi})}]
  \item	Automatically downloading a large amount of papers from academic libraries is generally restricted to registered memberships, while many public Web search engines offer APIs to enable retrieving search results programmatically (e.g., Google's JSON/Atom Custom Search API).
  \item	SLR implementations normally yield a relatively low amount of primary studies, Web content mining could involve a huge quantity of Web pages, which makes it difficult to manually screen the Web content to be mined.
  \item	The knowledge in academic publications are generally comprehensive and complicated, which inevitably leads to labor-intensive efforts on the data collection and synthesis in an SLR. On the contrary, the information in technical websites and forums are usually easy to understand at the introductory or intermediate levels in plain texts, and thus the corresponding data identification and analysis are suitable for automation in this case (e.g., employing suitable text data mining techniques).
\end{enumerate} 

Therefore, we propose to adapt the SLR methodology and make it align with the characteristics of Web content mining and knowledge discovery.

The remainder of this paper is organized as follows. 
Section \ref{sec:adaptation} specifies our proposal about adapting the methodology SLR to Web content mining. In particular, our current focus is mainly on adapting the stage of Conducting Review. Conclusions and some future work are discussed in Section \ref{sec:conclusion}.

\section{Methodology Adaptation}
\label{sec:adaptation}
The proposed methodology aims to generally explain and guides the whole research practices so that it is applicable to individual evidence-based research actives by means of specific tools,  techniques and processes of Web mining and knowledge discovery. Hence, we adapt the SLR methodology to Web mining and knowledge discovery along two directions. On one hand, we align the guidelines of SLR implementations with Web mining activities; on the other hand, we integrate suitable Web mining methods to the SLR methodology. In particular, we mainly focus on the stage of Conducting Review, and propose a mapping of the activities between systematic review and Web mining, as shown in Table \ref{tbl>adaptation}. 

Note that the Web mining activities Programmatic Searching and Supervised Topic Modeling are together responsible for primary study selection.  

\begin{table}[!t]\footnotesize
\renewcommand{\arraystretch}{1.4}
\centering
\caption{\label{tbl>adaptation}Adaptation Mapping for the Stage of Conducting Review}
\begin{tabular}{|>{\raggedright}p{2.2cm}|>{\raggedright\arraybackslash}p{5.2cm}|}
\hline
\textbf{Web Mining} & \textbf{Systematic Review} \\
\hline
\multirow{4}{2.2cm}{Programmatic Searching} & Exhaustively search relevant primary studies in the literature.\\ 
\cline{2-2}
& \multirow{3}{5.2cm}{Select relevant primary studies and assess their qualities for answering research questions.}\\

 & \\
\cline{1-1}
\multirow{3}{2.2cm}{Supervised Topic Modeling} & \\
\cline{2-2}
& Extract useful data from the selected primary studies.\\
\hline
Text Categorization \& Document Clustering & \multirow{5}{5.2cm}{Arrange and synthesize the initial review results into review notes.}\\
\cline{1-1}
Association Rule Mining & \\
\hline
\end{tabular}
\end{table}

\subsection{Programmatic Searching}
It has been identified that the rigor of the search process is one of the distinctive characteristics of systematic reviews \cite{Zhang_Babar_2010}. The adaptation here is to make the search process programmatic so as to reduce possible human bias and improve the rigor of the searching. The search process in traditional SLR implementations is of manual actives due to the policies of academic libraries. Search strings are manually input to the literature search engines, and further download candidate publications individually from the search results. 

When it comes to the Web mining, as the name suggests, useful data are located in the public Web and there is usually no limit for data source saving from webpages. As such, we consider to improve SLR's search activity to be a programmatic process that is composed of two steps. Firstly, we programmatically retrieve Web search results by calling the public search engine APIs. For example, Google supplies the JSON/Atom Custom Search API\footnote{\url{https://developers.google.com/custom-search/json-api/v1/overview}} to allow RESTful requests to get Web search results in the JSON or Atom format; and Microsoft offers the Bing Search API\footnote{\url{https://datamarket.azure.com/dataset/5BA839F1-12CE-4CCE-BF57-A49D98D29A44}} to enable collecting Web search results using the XML or JSON format. In particular, it is possible to enlarge search scope to multiple websites by choosing the same format of results. The search results are records comprising data source addresses, i.e.~the Uniform Resource Locators (URLs). 
Secondly, once the URLs are retrieved, it is convenient to develop programs to obtain and save the content of the corresponding webpages in a structure of interest.

It is notable that the programmatic searching is not supposed to replace necessary manual searching especially in Pilot Review. In fact, according to the aforementioned procedure of implementing an SLR (cf.~Figure \ref{fig_SLR_procedure}), the programmatic searching belongs to Conducting Review, while Planning Review largely relies on the manual searching-based pilot review that can help gradually improve search strategy, refine inclusion/exclusion criteria, and verify data extraction schema. In other words, the manual searching can be viewed as a prerequisite for the programmatic searching. For example, we can follow the Quasi-Gold Standard (QGS) based manual search strategy \cite{Zhang_Babar_2010} to determine the most suitable programmatic search string, according to its search performance in terms of sensitivity and precision. Moreover, the refined inclusion/exclusion criteria can also be integrated into codes to facilitate programmatic searching.

\subsection{Supervised Topic Modeling}
Given the predefined research questions, traditional SLRs use a data extraction schema to collect relevant data from primary studies. The schema covers a set of attributes, and each attribute corresponds to a data item. For a selected study, the data items usually have to be identified by screening and understanding the academic descriptions. Considering that the online information is generally straightforward, the adaptation here is to use an information retrieval technique, namely supervised topic modeling to improve the study selection and meanwhile conduct the raw data extraction.

In theory, topic modeling can automatically find overarching topics from a given text corpus, without predefining tags, taxonomies, or training data \cite{Blei_Lafferty_2009}.   Recall that topic modeling is based on the word frequencies and co-occurrence frequencies in the relevant documents. In this case, we propose to use the predefined data extraction schema to ``supervise" the topic modeling activities. Given the specific schema attributes,  we first use the attribute word frequencies to choose relevant Web posts/pages from the searched candidates; secondly, we keep the context information as the raw data when building a model of related words. 

\begin{figure}[!t]
\centering
\includegraphics{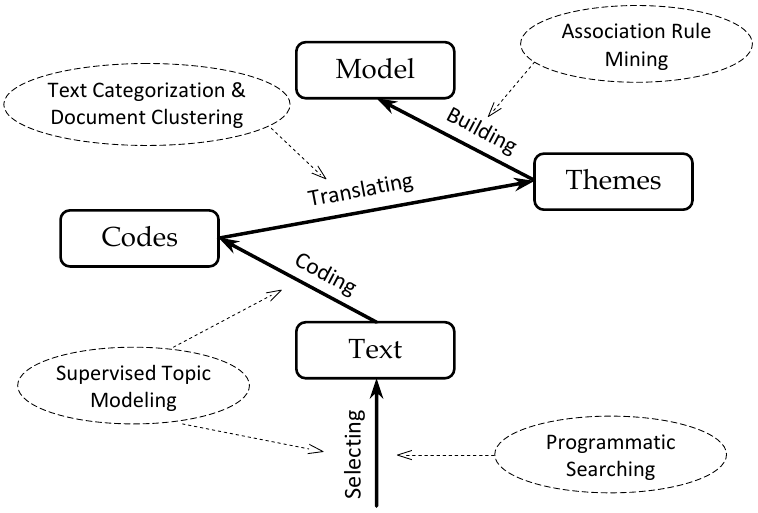}
\caption{Data evolution and its corresponding Web mining activities during thematic synthesis and aggregation.}
\label{fig_data_form}
\end{figure}

 In fact, the main purpose of employing Supervised Topic Modeling is to split the selected Web content into pieces, and then settle them into suitable columns of the data extraction schema on a webpage-by-webpage basis.  Since the raw data are mostly qualitative descriptions, the whole data synthesis and aggregation are supposed to follow the approach of thematic analysis \cite{Cruzes_2011}. As suggested by Cruzes and Dyb{\aa} \cite{Cruzes_2011}, the process of thematic synthesis
and aggregation drives different forms of the data with
an increasing level of abstraction, as shown in Figure \ref{fig_data_form}. The four
types of data forms are briefly explained below.

\begin{itemize*}
    \item	\textbf{Text} refers to the raw data with qualitative descriptions.
    \item	\textbf{Codes} are descriptive labels that represent different segments of the raw data.
    \item	\textbf{Themes} categorize the initial codes into a smaller set of concentrated-meaning units.
    \item	\textbf{Model} denotes taxonomy or theory that portrays a big picture consisting of higher-order themes and their relationships.
\end{itemize*}

Since the essential activities of this step are to identify and deal with keywords and key-phrases, we still consider the data form here to be ``Text" and ``Codes" in the thematic analysis process, although the generated model of topics can be viewed as initial mapping-study analysis results.

\subsection{Replaceable Text Mining Techniques for Data synthesis}
\label{subsec:textMining}

 Although there are various information types online, text is still the most commonly used type of unstructured Web information \cite{Grimes_2008,Mitra_2016}. Therefore, we try to reuse the existing text mining techniques for data synthesis in our methodology. As a demonstration, we focus on text categorization, document clustering, and association rule mining in this paper. In practice, different problems would have to be solved using different approaches, and thus practitioners could replace these techniques with suitable alternatives.

\subsubsection{Text Categorization and Document Clustering}
Once the raw data are collected and arranged in the aforementioned schema columns, we propose to employ the techniques text categorization and document clustering to make the data form evolve into Themes, column by column.
Text categorization is still a ``supervised" technique for the situation where the categories are known beforehand. Take the investigation in Cloud API issues \cite{Lu_Zhu_2013} as an example, suppose there is a data column of Cloud APIs, we can naturally use the available APIs to categorize the selected webpages. In contrast, document clustering is
an ``unsupervised" method for grouping documents/text without predefining categories or classes \cite{Chebel_Latiri_2015}. Suppose the Cloud API issue investigation also has a data column of API issues, we will have to use document clustering to identify the groups of webpages that belong together, because it is impossible for us to pre-understand what API issues could happen. In the study \cite{Lu_Zhu_2013}, for example, the four major types of Cloud API issues are halt failures, content failures, late timing failures, and erratic failures.

\subsubsection{Association Rule Mining}
After categorizing each schema column's raw data into a smaller set of concentrated-meaning units, we propose to employ the technique association rule mining to generate higher-order knowledge Models. Association rule mining is a popular method for discovering frequent patterns, correlations, or causal structures among variables in large databases. Here we imagine the data extraction schema as a database that is composed of the collected raw data. Unlike text categorization and document clustering that group a single column's data items into Themes, association rule mining is to identify common Themes that co-occur frequently across different columns. For example, suppose there are other data columns such as conditions, configurations and human activities in the data schema of the study on Cloud API issues \cite{Lu_Zhu_2013}; then we will be able to use association rule mining to automatically find potential antecedent-consequent relations such as the trigers for different API issues.

\section{Conclusions}
\label{sec:conclusion}
In addition to the academic literature, the public and technical information in the Web is becoming an alternative type of evidence for EBSE studies. Unlike traditional EBSE studies that widely employ the rigorous SLR, it is a lack of a standard and systematic methodology for mining Web content as evidence. We propose to apply the methodology SLR to Web content mining and knowledge discovery in software engineering. The direct application of SLR is not feasible due to the significant differences between reviewing academic literature and investigating Web knowledge. As such, we decide to adapt SLR to the characteristics of Web content by generally following the logic of SLR while adjusting detailed activities, in particular, making the stage of Conducting Review (cf.~Figure \ref{fig_SLR_procedure}) automated with Web mining techniques. Therefore, we outline an adaptation mapping between Web mining and systematic review mainly for this stage. Our future work is then to develop a concrete prototype of the adapted methodology with implementation of the automated components ranging from programmatic searching to various replaceable text mining techniques.


\section{Acknowledgments}
This work is supported by the Swedish Research Council
(VR) for the project ``Cloud Control", and through the
LCCC Linnaeus and ELLIIT Excellence Centers.

%
\bibliographystyle{abbrv}
\bibliography{sigproc}  
%
%
\end{document}